\documentclass[aps,prl,preprint,superscriptaddress]{revtex4-2}

\usepackage{graphicx}
\usepackage{dcolumn}
\usepackage{amsmath,amssymb,amsfonts,bm}
\usepackage{booktabs}
\usepackage{mathrsfs}

\newcommand{\ket}[1]{\lvert #1\rangle}
\newcommand{\bra}[1]{\langle #1\rvert}
\newcommand{\braket}[2]{\langle #1\mid #2\rangle}

\begin{document}

\preprint{A journal version of this work is under peer review}

\title{Quantum-Limited Symbol-Blind Channel Estimation for Coherent State Discrimination}

\author{Hongxu Chen}
\affiliation{State Key Laboratory of Networking and Switching Technology,
Beijing University of Posts and Telecommunications, Beijing 100876, China}
\affiliation{Beijing Key Laboratory of Convergent Communications and
Networking Technologies in LEO Satellite Systems, Beijing 100876, China}

\author{Renzhi Yuan}
\email[Contact author: ]{renzhi.yuan@bupt.edu.cn}
\affiliation{State Key Laboratory of Networking and Switching Technology,
Beijing University of Posts and Telecommunications, Beijing 100876, China}
\affiliation{Beijing Key Laboratory of Convergent Communications and
Networking Technologies in LEO Satellite Systems, Beijing 100876, China}

\author{Haifeng Yao}
\affiliation{School of Optics and Photonics,
Beijing Institute of Technology, Beijing 100081, China}

\author{Mugen Peng}
\affiliation{State Key Laboratory of Networking and Switching Technology,
Beijing University of Posts and Telecommunications, Beijing 100876, China}
\affiliation{Beijing Key Laboratory of Convergent Communications and
Networking Technologies in LEO Satellite Systems, Beijing 100876, China}

\date{\today}
\begin{abstract}
Residual dispersion breaks temporal-mode matching in photon-starved coherent links.  
For equiprobable $M$-ary PSK coherent states in a known spectral mode, with unknown symbols and carrier phase, we establish the quantum limit for blind joint estimation of group delay and second-order dispersion: after eliminating the common phase, it is  $4N_s\bm C$, set by the covariance of the centered generators alone. A multi-output quantum pulse gate with photon-number-resolving detection locally attains it and supports reception below the standard quantum limit under turbulent fading.

\end{abstract}
\maketitle
Quantum-enhanced detection enables discrimination of nonorthogonal quantum states
\cite{cook2007optical,becerra2013experimental,becerra2015photon}.
Coherent states, easy to prepare and loss-robust
\cite{giovannetti2004classical,weedbrook2012gaussian}, serve as natural carriers. 
Such receivers commonly implement optical displacement via the signal\cite{kennedy1973near,yuan2020free,bai2026near}.
Laboratory demonstrations typically derive both fields from a single laser and assume the channel preserves their mode matching. In long-distance links, unknown channel-induced waveform distortion breaks this assumption, and the resulting mismatch manifests as loss. 
Optical signals occupy finite-duration wave packets with structured spectra
\cite{raymer2020temporal,brecht2015photon},
so propagation can distort the received temporal mode. Since displacement detection projects the incident field onto the local-oscillator mode, accurate mode matching is essential 
\cite{blow1990continuum}.
Any mismatch reduces modal overlap and acts as an effective loss 
\cite{ferdinand2017multi,becerra2013experimental,becerra2015photon}.

Quantum-enhanced detection is increasingly applied to photon-starved deep-space and long-haul fiber links, 
where its advantages are most pronounced 
\cite{miao2026quantum,kakarla2020one}.
In free-space links, atmospheric turbulence induces intensity scintillation, phase fluctuations, and wavefront distortion. To mitigate these effects, picosecond or femtosecond pulses with dynamic coherent matched filtering can improve link stability, but their short duration heightens sensitivity to dispersion. Pulse broadening and temporal-mode mismatch then become major impairments.  
\cite{wang2021temporal,giorgetta2013optical,weiss2016symbol,wohlgemuth2024robust}.
Unlike free-space links, dispersion-induced waveform distortion remains a primary limitation on the reach of optical-fiber links 
\cite{chen2023continuous,kakarla2020one}.
Existing optical compensation uses static or tunable devices, whose limited adaptability leaves residual dispersion uncompensated
\cite{gruner2005dispersion}.
Coherent detection with digital signal processing offers  an electrical-domain alternative but requires a more complex receiver and a strong local-oscillator, both imposing  practical limitations.

In this paper, we propose a method for estimating slowly varying residual group-delay dispersion (GDD) after coarse compensation
\cite{kato2000temperature}.
Small positive or negative dispersions yield nearly indistinguishable temporal intensity profiles, which cause direct detection to lose first-order sensitivity. In the weak-field regime, the information available to standard heterodyne detection decreases quadratically with the signal photon number. We derive the precision limit of dispersion estimation in photon-starved links  in the form of a quantum Cram\'er--Rao bound (QCRB)
\cite{helstrom1969quantum,tsang2016quantum}.
Remarkably, after eliminating the unknown carrier phase, the effective quantum Fisher information (QFI) matrix takes a universal form, preserving parameter information. We construct an experimentally implementable positive operator valued measure (POVM): a quantum pulse gate (QPG) projects the field onto phase-diverse temporal modes, and photon-number-resolving detectors (PNRDs) record output counts. This measurement extracts all effective quantum information and locally attains the QCRB. Count differences retain both magnitude and sign of residual GDD. The estimates also recover the displacement local-oscillator waveform. 
We compare its performance with the balanced-quadrature reconstruction model of Morin \textit{et al.}
\cite{morin2013experimentally,morin2020accurate}.
Finally, we combine the estimator with multidimensional temporal-mode encoding.
The resulting architecture supports reception below the standard quantum limit (SQL) without instantaneous turbulence estimates.

For propagation in a fixed beam geometry or a waveguide such as an optical fiber, the field may be treated as being in a single spatial mode 
\cite{blow1990continuum,brecht2015photon,raymer2020temporal}.
High-speed modulation and channel dispersion produce structured temporal wave packets, requiring a continuous-mode description.
Let $\hat a(\omega)$ denote the frequency-resolved field operator, with
$[\hat a(\omega),\hat a^\dagger(\omega')]=\delta(\omega-\omega')$.
A normalized wave packet $q_{\bm\vartheta}(\omega)$ defines the temporal-mode operator
$
\hat a[q_{\bm\vartheta}]
=
\int d\omega\,
q_{\bm\vartheta}^{*}(\omega)\hat a(\omega).
$
Thus, $\ket{\alpha,q_{\bm\vartheta}}$ denotes a coherent state obtained by displacing this temporal mode by the complex amplitude $\alpha $.

The transmitted mode and channel jointly determine the received wave packet, with dispersion a primary source of mode mismatch in short-pulse links 
\cite{chen2023continuous,wang2021temporal}.
Neglecting third- and higher-order dispersion, we define
$\Omega=(\omega-\omega_0)/(\sqrt{2}\sigma_\omega)$
and let $q(\Omega)$ denote the normalized transmitted spectrum, with the Jacobian absorbed.
For $\bm\theta=(\varphi_c,\bm\vartheta)$ and $\bm\vartheta=(\tau,\kappa)$,
\[
q_{\bm\vartheta}(\Omega)
=
e^{i(\tau\Omega+\kappa\Omega^2/2)}q(\Omega),
\]
where $\varphi_c$ is carried by the coherent amplitude, $\tau$ shifts the arrival time, and $\kappa$ produces broadening and chirp.

If the mean received photon number per symbol is $N_s$, the photon count in $[t,t+dt]$ is a Poisson random variable with mean
$N_s\Lambda_{\bm\vartheta}(t)dt$
\cite{shapiro2009quantum}.
Since $N_s$ is independent of the unknown parameters, the classical FI from direct detection is
\begin{equation}
    J_{\mu\nu}^{\mathrm{dir}}
    =
    N_s\int_{-\infty}^{\infty}
    \frac{
    \partial_{\vartheta_\mu}\Lambda_{\bm\vartheta}(t)\,
    \partial_{\vartheta_\nu}\Lambda_{\bm\vartheta}(t)}
    {\Lambda_{\bm\vartheta}(t)}
    \,dt
    \label{eq:direct-fisher}.
\end{equation}
where $\Lambda_{\bm\vartheta}(t)=|\tilde q_{\bm\vartheta}(t)|^2$

We consider equiprobable BPSK signals occupying a single temporal mode $q_{\bm\vartheta}$ per symbol interval.
Each symbol contains a mean received photon number $N_s$, while the common carrier phase $\varphi_c$ is unknown.
In the single-photon frequency space, group delay $\tau$ and second-order dispersion $\kappa$ are generated by the Hermitian operators
$\hat g_\tau=\hat\Omega$ and $\hat g_\kappa=\hat\Omega^2/2$.
Each mode derivative contains a component parallel to the original temporal mode.
This component changes only the global phase and is therefore indistinguishable from the unknown common carrier phase.
After projecting it out , the covariance matrix $\bm C$ of the centered generators determines the first-order observable sensitivities and their coupling
\begin{equation}
    \bm C
    =
    \begin{pmatrix}
        \langle\Omega^2\rangle-\langle\Omega\rangle^2
        &
        \dfrac{1}{2}
        \left(
            \langle\Omega^3\rangle
            -
            \langle\Omega\rangle
            \langle\Omega^2\rangle
        \right)
        \\[3mm]
        \dfrac{1}{2}
        \left(
            \langle\Omega^3\rangle
            -
            \langle\Omega\rangle
            \langle\Omega^2\rangle
        \right)
        &
        \dfrac{1}{4}
        \left(
            \langle\Omega^4\rangle
            -
            \langle\Omega^2\rangle^2
        \right)
    \end{pmatrix}
    \label{eq:main-generator-covariance}
\end{equation}
Its entries are functions of the raw spectral moments
$
\langle\Omega^n\rangle_q
\equiv
\langle q|\hat\Omega^n|q\rangle
=
\int d\Omega\,\Omega^n|q(\Omega)|^2
$
of the normalized spectral intensity $|q(\Omega)|^2$. 

The blind BPSK state is a statistical mixture of two nonorthogonal coherent states in an infinite-dimensional Fock space. Direct evaluation of its QFI matrix requires solving the infinite-dimensional SLD Lyapunov equation
\cite{liu2020quantum,tsang2016quantum}.
We exploit the correspondence between a BPSK sign flip and total photon-number parity to decompose the mixed state exactly into two orthogonal parity sectors and obtain a closed-form QFI.
With the even- and odd-parity projectors
\cite{dodonov1974even}, the state becomes
\begin{equation}
    \rho_{\bm\theta}
    =
    \lambda_e
    \ket{\mathrm{cat}_e(\bm\theta)}
    \bra{\mathrm{cat}_e(\bm\theta)}
    +
    \lambda_o
    \ket{\mathrm{cat}_o(\bm\theta)}
    \bra{\mathrm{cat}_o(\bm\theta)},
    \label{eq:parity-resolved-decomposition}
\end{equation}
where $\lambda_{e/o}=(1\pm e^{-2N_s})/2$ and 
$
\ket{\mathrm{cat}_{e/o}}\propto
\ket{\sqrt{N_s}e^{i\varphi_c},q_{\bm\vartheta}}\pm
\ket{-\sqrt{N_s}e^{i\varphi_c},q_{\bm\vartheta}}
$ 
are the normalized even/odd cat states \cite{dodonov1974even}.

The common phase, delay, and dispersion are generated by photon-number-preserving unitaries and therefore preserve parity
\cite{brecht2015photon,raymer2020temporal}.
Since $\lambda_p$ depends only on $N_s$ and
$\bra{\mathrm{cat}_e}\hat H_\mu\ket{\mathrm{cat}_o}=0$,
the mixed-state QFI reduces exactly to the $\lambda_p$-weighted sum of the pure-sector generator covariances.
To avoid differentiating infinitely many Fock components directly, we introduce the commuting Hermitian generators $\hat H_\mu$, for which
$
    \partial_\mu
    \ket{\mathrm{cat}_p(\bm\theta)}
    =
    i\hat H_\mu
    \ket{\mathrm{cat}_p(\bm\theta)}.
$
The QFI matrix is then
\begin{equation}
    \mathcal K_{\mu\nu}
    =
    4\sum_{p=e,o}
    \lambda_p
    \operatorname{Re}
    \left[
        \langle\hat H_\mu\hat H_\nu\rangle_p
        -
        \langle\hat H_\mu\rangle_p
        \langle\hat H_\nu\rangle_p
    \right].
    \label{eq:total-generator-qfim}
\end{equation}
The generators are the second quantizations $\hat H_\mu=\int d\Omega\,g_\mu(\Omega)\,\hat a^\dagger(\Omega)\hat a(\Omega)$, with $\hat H_{\varphi_c}=\hat N$.
Eq.~\eqref{eq:total-generator-qfim} converts the infinite-dimensional SLD problem into generator-covariance calculations in the even and odd cat states.
Because the Fock-space generators are second quantizations of single-photon spectral generators, their first and second moments are fixed by two sets of statistics: the spectral moments of the single-photon temporal mode and the photon-number moments in each parity sector.
For the normalized input spectrum $q(\Omega)$, group delay and second-order dispersion modify only the spectral phase, leaving the spectral intensity unchanged:
$
|q_{\bm\vartheta}(\Omega)|^2=|q(\Omega)|^2.
$

The required photon-number moments are $\langle\hat N\rangle_e=N_s\tanh N_s$,
$\langle\hat N\rangle_o=N_s\coth N_s$, and
$\langle\hat N(\hat N-1)\rangle_{e/o}=N_s^2$.
The latter enter only through the positivity of the fluctuation term and drop out of the final result.
Substitution into Eq.~\eqref{eq:total-generator-qfim} yielding a closed-form three-parameter QFI matrix.
Because the common carrier phase $\varphi_c$ is unknown, the effective QFI matrix is the Schur complement
$
    \bm{\mathcal K}_{\bm\vartheta|\varphi_c}^{\mathrm{eff}}
    =
    \bm{\mathcal K}_{\bm\vartheta\bm\vartheta}
    -
    \bm{\mathcal K}_{\bm\vartheta\varphi_c}
    \mathcal K_{\varphi_c\varphi_c}^{-1}
    \bm{\mathcal K}_{\varphi_c\bm\vartheta}.
$

Within each parity sector, the generator covariance splits into mode-deformation and photon-number-fluctuation parts,
$
\operatorname{Cov}_p(\hat H_\mu,\hat H_\nu)
=\langle\hat N\rangle_p\,C_{\mu\nu}
+\operatorname{Var}_p(\hat N)\,\bar g_\mu\bar g_\nu ,
$
where $\bar g_\mu=\langle g_\mu\rangle_q$ and $g_{\varphi_c}\equiv1$.
Because the sectors are orthogonal with parameter-independent weights, the
blocks of the full QFI matrix are the $\lambda_p$-weighted sums,
$
\mathcal K_{\mu\nu}=4\big[N_sC_{\mu\nu}+\bar V\,\bar g_\mu\bar g_\nu\big],
$
with $N_s=\sum_p\lambda_p\langle\hat N\rangle_p$ and
$\bar V=\sum_p\lambda_p\operatorname{Var}_p(\hat N)>0$ the mean within-sector
photon-number variance. Since $g_{\varphi_c}$ is constant, $C_{\varphi_c\mu}=0$,
so the rank-one fluctuation term carries the entire $\varphi_c$ row and column
and is canceled identically by the Schur complement, leaving
$
    \bm{\mathcal K}_{\bm\vartheta|\varphi_c}^{\mathrm{eff}}
    =
    4N_s\bm C .
$
The same construction applies to any $M$-PSK constellation, whose blind mixture
decomposes into $M$ orthogonal photon-number sectors mod $M$, the generators
preserve each sector and the effective QFI is again $4N_s\bm C$ (see Appendix Sec.~\ref{em:multimode} for the multimode case).

Define the Gaussian sign reliability function
$
H(\xi)
\equiv
\mathbb E_X\!\left[
\tanh^2\!\left(2\sqrt{\xi}\,X\right)
\right],
X|s\sim\mathcal N\!\left(s\sqrt{\xi},\frac12\right),
$
where $s=\pm1$ with equal probability and the expectation is taken over the resulting Gaussian mixture.
Eliminating the unknown carrier phase $\varphi_c$ by a Schur complement
\cite{suzuki2020quantum}, the Gaussian Fisher-information matrices are
\begin{equation}
\bm J_{\bm\vartheta|\varphi_c}^{G,\max}
=
4N_sH(2N_s)\bm C,
\qquad
\bm J_{\bm\vartheta|\varphi_c}^{\mathrm{het}}
=
2N_sH(N_s)\bm C.
\label{eq:gaussian-fi-comparison}
\end{equation}
The Gaussian optimum (see Appendix, Sec.~\ref{em:gaussian}) is obtained by homodyning the carrier amplitude quadrature to infer the unknown BPSK sign while measuring the phase quadratures of the orthogonal score modes.
These observables commute, as they occupy distinct temporal modes.
Heterodyne detection probes the same covariance matrix $\bm C$, but additional vacuum noise reduces sign reliability.

We construct a mode-selective estimator based on QPG and PNRD detection that attains the QCRB.
A QPG performs a coherent, mode-selective frequency conversion singling out  one temporal mode while leaving orthogonal modes untouched
\cite{brecht2015photon, brecht2014demonstration, donohue2018quantum}, 
and demonstrated that multi-output QPGs (MQPG) project simultaneously onto a programmable temporal-mode basis mapped to separate frequency channels \cite{serino2023realization}.

The construction follows from the observation that all usable first-order information lies in the local tangent space orthogonal to the operating-point mode, as the parallel component is indistinguishable from the common phase. We span this space with two score modes and mix them with the operating-point mode as an internal phase reference, converting phase deformations into linear PNRD count changes. 
After coarse compensation, fix
$\bm\vartheta_0=(\tau_0,\kappa_0)$ and write
$\ket{q_0}\equiv\ket{q_{\bm\vartheta_0}}$.
Since the parameters $\vartheta_\mu$, $\mu\in\{\tau,\kappa\}$, enter only through
the spectral phase,
$\partial_\mu\ket{q_{\bm\vartheta}}=ig_\mu\ket{q_{\bm\vartheta}}$.
With $\bar g_\mu=\langle g_\mu\rangle_{q_0}$, define
$\ket{\partial_\mu^\perp q}
=(I-\ket{q_0}\bra{q_0})
\partial_\mu\ket{q_{\bm\vartheta}}|_{\bm\vartheta_0}
=i(g_\mu-\bar g_\mu)\ket{q_0}$.
The remaining component,
$i\bar g_\mu\ket{q_0}$, changes only the common phase and is removed with the unknown carrier phase.
The projected derivatives satisfy
$\operatorname{Re}\braket{\partial_\mu^\perp q}
{\partial_\nu^\perp q}=C_{\mu\nu}$
and span a subspace $\mathcal D$ of dimension
$r=\operatorname{rank}\bm C\leq2$.

Let $\ket{v_\mu}\equiv-i\ket{\partial_\mu^\perp q}$.
Because $\braket{v_\mu}{v_\nu}=C_{\mu\nu}$ is real, an orthonormal basis
$\{\ket{e_a}\}_{a=1}^{r}$ of $\mathcal D$ can be chosen such that
$\ket{v_\mu}=\sum_aR_{a\mu}\ket{e_a}$,
where $R$ is real and $\bm C=R^{\!\top}R$.
The received mode then has the local expansion
\begin{equation}
\ket{q_{\bm\vartheta_0+\delta\bm\vartheta}}
=
e^{i\chi(\delta\bm\vartheta)}
\left[
\ket{q_0}
+i\sum_{a,\mu}R_{a\mu}\delta\vartheta_\mu\ket{e_a}
\right]
+\mathcal O(\|\delta\bm\vartheta\|^2),
\label{eq:local-mode-expansion}
\end{equation}
where
$\chi(\delta\bm\vartheta)=
\sum_\mu\bar g_\mu\delta\vartheta_\mu$
to first order.
Thus, the signed perturbations are encoded in the phase quadratures of at most two modes.
Their direct counts are quadratically dependent on the perturbations, sign-insensitive, and locally non-regular, whereas coherent mixing with $\ket{q_0}$ produces linear count differences.
Local estimation therefore can be performed using at most three temporal modes.

\begin{figure}[t]
  \centering
  \includegraphics[
    width=\columnwidth,
    trim={0cm 10.5cm 0cm 10.5cm},
    clip
  ]{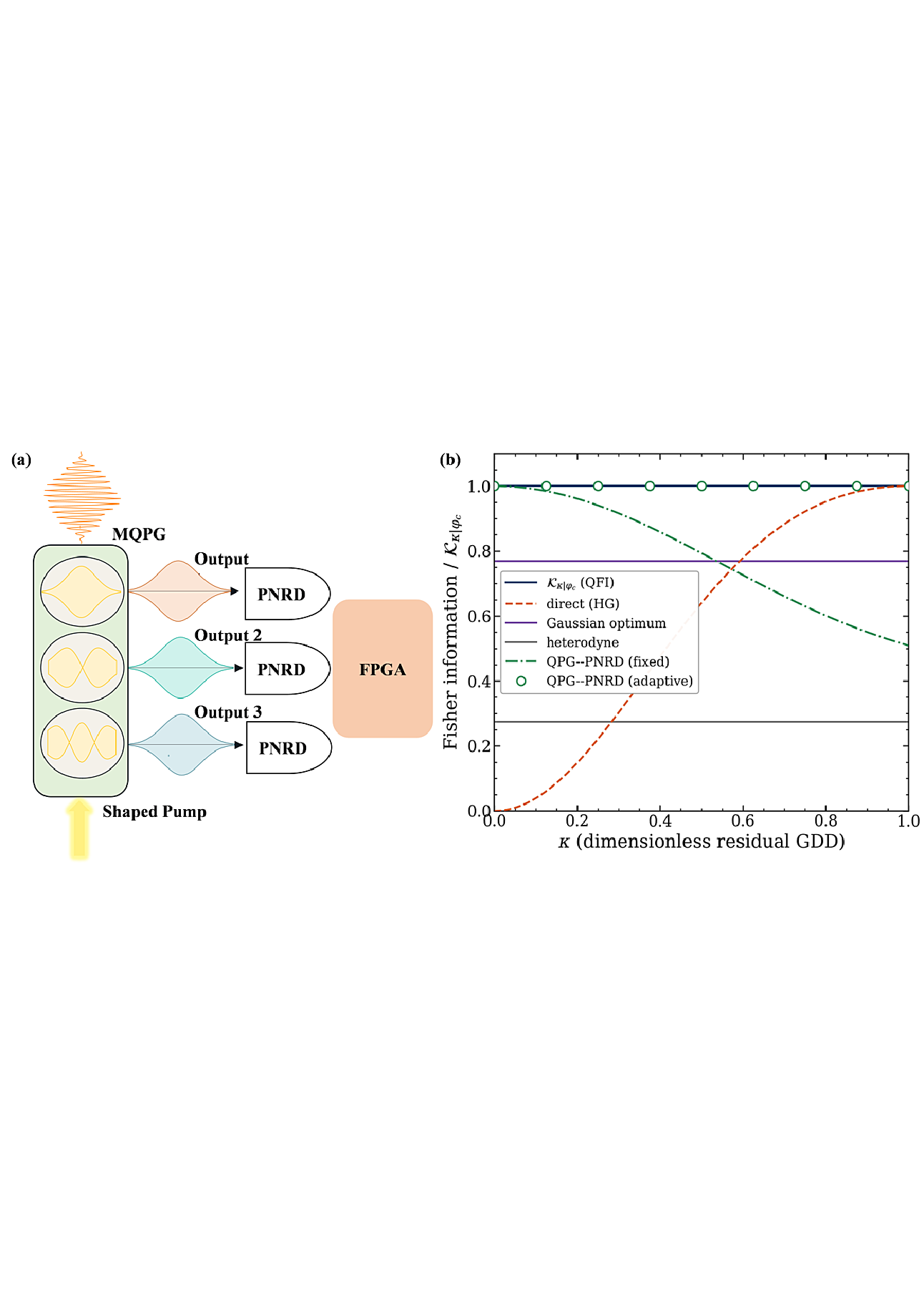}
  \caption{
  (a) An MQPG implements the three-mode analysis and routes the outputs to frequency-resolved PNRDs.
  (b) Fisher information for the dimensionless residual GDD $\kappa$, normalized by the effective QFI.}
  \label{fig1:main}
\end{figure}

For the generic rank-two case, an ideal unit-efficiency MQPG maps
$\{\ket{q_0},\ket{e_1},\ket{e_2}\}$
onto three resolvable frequency channels
\cite{serino2023realization} through
\begin{equation}
U^\star
=
\frac{1}{\sqrt{3}}
\begin{pmatrix}
1 & i\sqrt{3/2}  & i/\sqrt{2} \\
1 & -i\sqrt{3/2} & i/\sqrt{2} \\
1 & 0             & -i\sqrt{2}
\end{pmatrix},
\qquad
U^{\star\dagger}U^\star=I_3 .
\label{eq:optimal-three-mode-unitary}
\end{equation}
The analysis mode at output $j$ is
$\ket{m_j}
=(U^\star)_{j1}^{*}\ket{q_0}
+(U^\star)_{j2}^{*}\ket{e_1}
+(U^\star)_{j3}^{*}\ket{e_2}$.
As shown in Fig.~\ref{fig1:main}(a), the first column distributes
$\ket{q_0}$ equally among the outputs, while the remaining columns map the two score modes onto independent count-difference directions.
Their phases convert the imaginary score-mode amplitudes into linear intensity changes relative to $\ket{q_0}$.
For a higher-dimensional derivative subspace, $U^\star$ must be enlarged accordingly, but the construction is unchanged.

A lossless transformation maps the conditional coherent state to a product of coherent output states
\cite{loudon1974quantum,reck1994experimental}.
The PNRD counts are therefore independent Poisson variables with means
$\nu_j(\bm\vartheta)
=N_s|\braket{m_j}{q_{\bm\vartheta}}|^2$.
The unknown PSK symbol and carrier phase multiply all output amplitudes by a common phase and hence not affecting  the count distribution.
At the operating point,
$\nu_j(\bm\vartheta_0)=N_s/3$.
Substitution of Eq.~\eqref{eq:local-mode-expansion} into the Poisson FI gives
\begin{equation}
\bm J(U^\star)
=
4N_sR^{\!\top}R
=
4N_s\bm C
=
\bm{\mathcal K}_{\bm\vartheta|\varphi_c}^{\mathrm{eff}} .
\label{eq:optimal-fim-equals-qfim}
\end{equation}
The three-mode MQPG--PNRD measurement locally extracts the full effective QFI.
Uniform QPG conversion loss replaces $N_s$ by $\eta_{\rm QPG}N_s$ in Eq.~\eqref{eq:optimal-fim-equals-qfim}, while calibrated modal crosstalk only rescales $\bm J$ and preserves the unbiased count contrasts in Eq.~\eqref{eq:port-estimators}.

The quantum limit and its local attainability hold for any known single temporal mode. As a representative example, take the normalized fundamental Hermite–Gaussian (HG) mode. 
Its delay and dispersion score modes are $h_1$ and $h_2$, respectively.
Consequently,
$\mathcal K_{\tau\tau|\varphi_c}^{\mathrm{eff}}=2N_s$
and
$\mathcal K_{\kappa\kappa|\varphi_c}^{\mathrm{eff}}=N_s/2$,
independent of $\tau$ and $\kappa$.

We consider representative photon-starved communication scenarios, with typical residual dispersion in long-distance links after fixed coarse compensation. This residual dispersion arises from free-space path-length fluctuations and temperature variations in optical fibers 
\cite{chen2023continuous,wang2021temporal}.
Without timely compensation, it accumulates and substantially degrades temporal-mode matching.
Here $\kappa$ is dimensionless, and the calculated residual dispersion for the representative photon-starved links satisfies $|\kappa|\ll1$.
At the coarse-compensation point, direct detection has no local dispersion sensitivity, while Gaussian measurements remain below the quantum limit for finite $N_s$.
The adaptive QPG--PNRD receiver attains the QCRB, whereas the fixed receiver remains quantum limited to leading order for small residual dispersion, avoiding adaptive compensation in this regime.
For this mode, $\tau$ only translates the Gaussian intensity, so direct detection attains $\mathcal K^{\rm eff}_{\tau\tau}$.
By contrast, the width scales as $\sqrt{1+\kappa^2}$, causing the direct-detection FI for $\kappa$ to vanish at coarse compensation Fig.~\ref{fig1:main}(b).
The QPG advantage therefore lies entirely in dispersion estimation.

Displacement-based quantum receivers require mode matching between the decision local oscillator and the received signal.
The mismatch
$1-|\braket{q_{\mathrm{LO}}}{q_{\bm\vartheta}}|^2$
acts as irreversible loss and can dominate error-probability degradation
\cite{becerra2013experimental,ferdinand2017multi,li2013suppressing}.
Recovering the LO is equivalent to identifying the received single-photon wave packet.  
Morin \textit{et al.} infer this mode nonparametrically by diagonalizing the autocorrelation kernel of continuous-wave homodyne records, with the leading eigenfunction giving the optimal waveform
\cite{morin2013experimentally}.
As a Gaussian measurement, its available information is bounded by
$\bm J_{\bm\vartheta|\varphi_c}^{G,\max}$.

For a dispersive channel, an analytical HG-basis parameterization recovers the local oscillator from two real coefficients rather than the full waveform.
Using $U^\star$, linear combinations of the PNRD port-count differences yield the locally unbiased estimators
\begin{equation}
\widehat{\delta\tau}
=
\frac{n_2-n_1}{2(n_1+n_2+n_3)}
\sqrt{\frac{3}{2}}\,
\frac{1}{\sigma_\omega},
\widehat{\delta\kappa}
=
\frac{2n_3-n_1-n_2}{2(n_1+n_2+n_3)}
\frac{1}{\sigma_\omega^2},
\label{eq:port-estimators}
\end{equation}
where $n_j$ is accumulated over $B$ symbols and
$\hat{\bm\vartheta}
=\bm\vartheta_0+\widehat{\delta\bm\vartheta}$.
$B$-symbol covariance attains $(4N_eB\bm C)^{-1}$.
The matched local oscillator, $\ket{\hat q}=\ket{q_{\hat{\bm\vartheta}}}$, can then be programmed with a pulse shaper.
For $\operatorname{rank}\bm C=2$, its mean mismatch is
\begin{equation}
\mathbb E[1-\mathcal F]_{\mathrm{QPG}}
\simeq
\operatorname{Tr}
\!\left[
\bm C\,\mathrm{Cov}(\hat{\bm\vartheta})
\right]
=
\frac{1}{2N_eB},
\label{eq:lo-fidelity}
\end{equation}
where $N_e$ is photons for estimation.
By comparison, the nonparametric Morin reconstruction gives 
$
\mathbb E[1-\mathcal F]_{\rm Morin}
\simeq N_{\rm mode}(1+1/N_e)/(N_eB)
$
for an equally partitioned complex mode
\cite{morin2020accurate}
, a mismatch larger by $2N_{\rm mode}(1+1/N_e)$ at the cost of a two-parameter channel model, valid after coarse compensation

\begin{figure}[t]
  \centering
  \includegraphics[
    width=\columnwidth,
    trim={0cm 7cm 0cm 6cm},
    clip
  ]{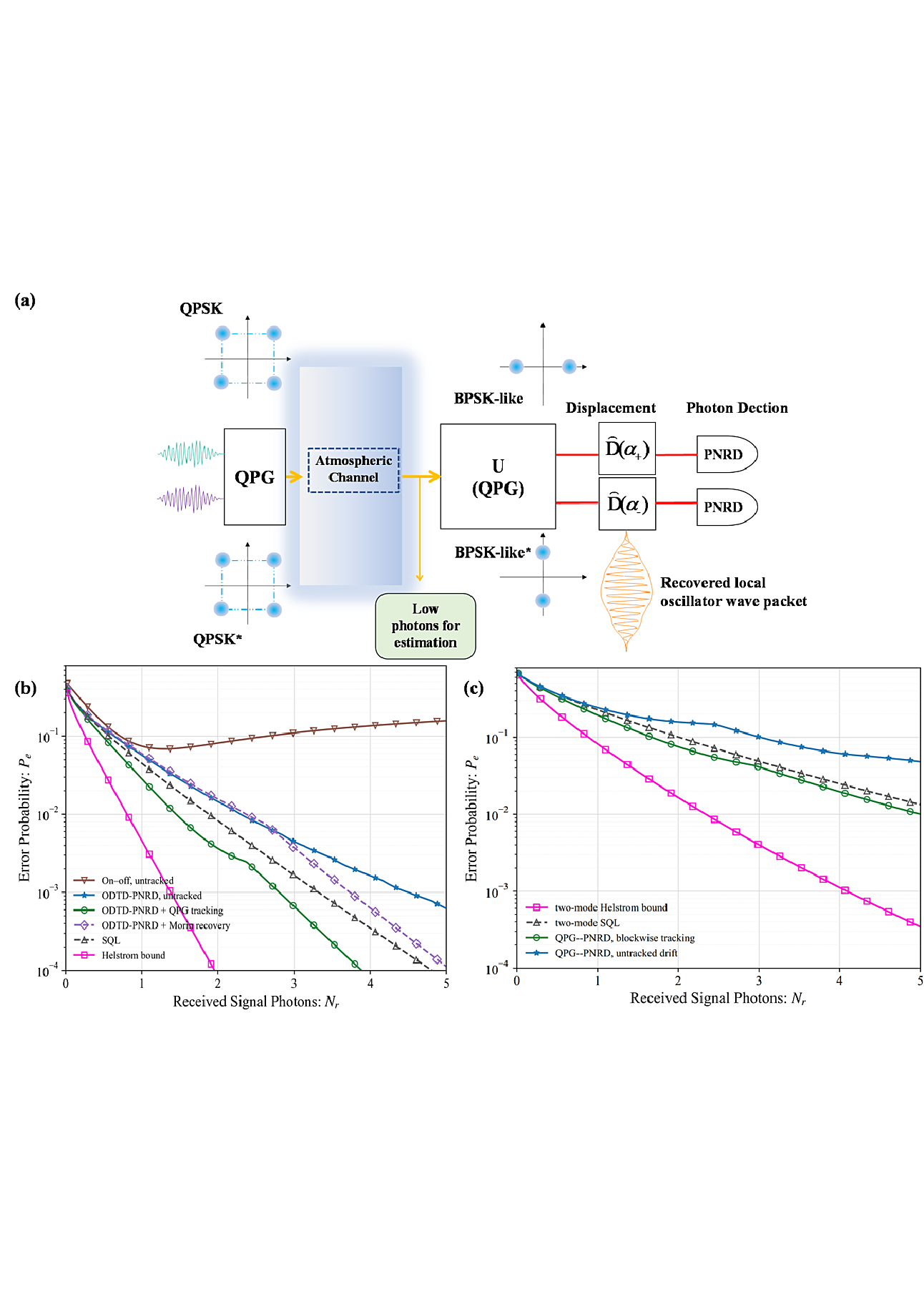}
  \caption{
    (a) Proposed temporal-mode-assisted receiver.
    (b) Error probabilities of Kennedy-type receivers with different mode-recovery strategies.
    (c) Error probability of the proposed two-mode QPSK receiver in turbulent fading.
  }
  \label{fig:main2}
\end{figure}

Multimode temporal encoding is an important application and provides additional freedom for receiver design 
\cite{brecht2015photon}.
We consider a robust QPSK receiver for a free-space channel with slowly varying dispersion and turbulence. The optimized displacement vector is designed offline. Marginalizing over the turbulence-transmittance prior avoids instantaneous estimation, while the residual-dispersion measurement retains the QFI Fig.~\ref{fig:main2}(a).
Even weak mode mismatch drives an on–off Kennedy receiver to an error floor as photon number increases, whereas optimally displaced threshold detection (ODTD) with a maximum-a-posteriori rule mitigates this saturation
\cite{yuan2020optimally,li2013suppressing}.
Fig.~\ref{fig:main2}(b) compares reception without recovery, with our parameter estimator, and with Morin-type waveform recovery under the same residual-dispersion model.
The kinks reflect decision-threshold changes. Only our scheme remains below the SQL throughout the simulated range and applies to most existing receivers.

Consider the two-temporal-mode QPSK codewords.
Both modes experience the same instantaneous intensity transmittance $\eta$\cite{cui2025superadditive}. Conditioned on $\eta$, the pure-loss channel maps
$\ket{\bm s_k}\mapsto\ket{\sqrt\eta\,\bm s_k}$.
The transmitter assigns equal energies and conjugate QPSK phases to the two modes, yields
$
\ket{\bm s_k}
=
\left|\sqrt{{N_s}/2}e^{i\theta_k}\right\rangle_{f_0}
\otimes
\left|\sqrt{{N_s}/2}e^{-i\theta_k}\right\rangle_{f_1}.
$
At the receiver, a QPG projects onto the selected modes and applies the Hadamard transformation
$
U_H=\frac{1}{\sqrt{2}}
\begin{pmatrix}
1&1\\
1&-1
\end{pmatrix},
f_\pm=\frac{f_0\pm f_1}{\sqrt{2}}.
$
The two outputs encode the signs of the in-phase and quadrature components, respectively.
Fixed displacements
$
\hat D(d_+)\otimes\hat D(i d_-)
$
are then applied along the corresponding decision axes, where
$d_\pm\geq0$.

The Hadamard transformation thus decomposes each QPSK symbol into two BPSK-like signs on orthogonal quadratures.
Defining
$b_{+,k}=\operatorname{sgn}(\cos\theta_k)$ and
$b_{-,k}=\operatorname{sgn}(\sin\theta_k)$,
the coherent mean photon numbers incident on the PNRDs are
$
\nu_{+,k}(\eta)
=
\eta N_s/2+d_+^2
+2b_{+,k}d_+\sqrt{\eta N_s/2},
$
and
$
\nu_{-,k}(\eta)
=
\eta N_s/2+d_-^2
+2b_{-,k}d_-\sqrt{\eta N_s/2}.
$
Turbulence changes the interference amplitudes through $\sqrt{\eta}$.
The fixed displacements therefore require no instantaneous channel-state information.

Both outputs are measured by PNRDs.
For detector efficiency $\eta_d$ and independent thermal background with mean photon number $N_t$ in each branch, the conditional count distribution is
$
p_{\rm th}(n\mid\nu)
=
\frac{(\eta_dN_t)^n}
     {(1+\eta_dN_t)^{n+1}}
\exp\!\left(
-\frac{\eta_d\nu}{1+\eta_dN_t}
\right)
L_n\!\left[
-\frac{\nu}{N_t(1+\eta_dN_t)}
\right],
$
where $L_n$ is the Laguerre polynomial
\cite{yuan2025multi}.
In the limit $N_t\rightarrow0$, this reduces to
$p_{\rm th}(n\mid\nu)
=e^{-\eta_d\nu}(\eta_d\nu)^n/n!$.
Dark counts are neglected.

The receiver applies maximum-a-posteriori (MAP) detection to the two counts
$\bm n=(n_+,n_-)$.
For codeword signs
$\bm b_k=(b_{+,k},b_{-,k})$
and prior $\pi_k$, the decision is
$
\hat k(\bm n)
=
\arg\max_k\,\pi_k\bar p(\bm n\mid k).
$
Conditioned on $\eta$, the branch counts are independent:
$
p(\bm n\mid\bm b_k,\eta)
=
\prod_{\ell\in\{+,-\}}
p_{\rm th}\!\left(n_\ell\mid\nu_{\ell,k}(\eta)\right).
$
Averaging over the log-normal transmittance gives
$
\bar p(\bm n\mid k)
=
\int_0^1d\eta\,p(\eta)\,
p(\bm n\mid\bm b_k,\eta).
$
The corresponding decision region is
$
\mathcal D_k
=
\{\bm n\in\mathbb Z_{\ge0}^2:
\pi_k\bar p(\bm n\mid k)
\ge
\pi_j\bar p(\bm n\mid j),\ \forall j\neq k\}.
$
The mean error probability is
\begin{equation}
P_e^{\mathrm{static}}
=
1-\sum_{\bm n\in\mathbb Z_{\ge0}^{2}}
\max_k\,\pi_k\bar p(\bm n\mid k).
\label{eq:static-error-probability}
\end{equation}
The fixed displacements are the only optimized parameters,
$
(d_+^\star,d_-^\star)
=
\arg\min_{d_+,d_-\ge0}
P_e^{\mathrm{static}}(d_+,d_-).
$
They are optimized offline over $p(\eta)$ and stored in a lookup table indexed by signal intensity, eliminating real-time channel estimation and feedback.

By symmetry, we can prove that the two optimized branch displacements are equal\cite{sidhu2021quantum}.
Since the estimator performs well by construction at vanishing residual dispersion, we instead probe its adaptability over a wider range.
The random residual-GDD change per estimation block is set to $0.5$.
We use $N_s=10$, $N_t = 0.001$,20 blocks, and turbulence variance of $\sigma_t^2=0.1$, only $N_e=3\%N_s$ photons are tapped for estimation, but $B=2000$ makes photon counts large, so thermal and dark noise are negligible.
Fig.~\ref{fig:main2}(c) shows that uncompensated dispersion substantially increases error probability, whereas our scheme remains below the SQL throughout the simulation (SQL and Helstrom benchmarks, Appendix Sec.~\ref{em:benchmarks}).
Without estimation, the fixed displacement is set at the nominal mismatch-free point, so its decision rule passes through a transition as the photon number grows, seen as the slope change of the untracked curve near $N_r\approx2.5$.
The tracking receiver re-centers each block and decreases smoothly and monotonically---precisely where blind estimation shows its advantage.

Using dispersion as a representative case, we establish the quantum limit for blind channel estimation in photon-starved links.  
A parity-sector decomposition yields the effective QFI matrix $4N_s\bm C$. 
We constructed a QPG--PNRD measurement that attains the QCRB and recovers dispersion information inaccessible to direct detection at the coarse-compensation point.
Even optimal Gaussian measurements retain only the fraction $H(2N_s)$ of this information. The same estimator recovers the matched LO without full waveform reconstruction. Since the port-count contrasts depend only on ratios, neither the estimator nor the decision rule requires instantaneous transmittance knowledge. Exploiting temporal-mode encoding, we designed a multimode QPSK receiver that combines parameter-based waveform recovery with unitary mode mixing. Simulations show error probabilities below the SQL under combined turbulence and residual dispersion. These results establish temporal-mode processing as a resource for quantum-enhanced detection in photon-starved long-distance links.

\begin{acknowledgments}
This work is supported by Beijing Natural Science Foundation under
Grant No.~4262010.
\end{acknowledgments}

\section*{Data Availability}
The data and simulation code supporting this study are available in the
Zenodo repository~\cite{chen2026dataset}.

\bibliography{ref}

\onecolumngrid
\appendix

\section{Optimal Gaussian measurement}\label{em:gaussian}
We prove that the multimode homodyne measurement in the main text
maximizes the effective FI among all Gaussian measurements.
The blind BPSK state has a nonnegative Wigner function
$W_{\rho_{\bm\theta}}(R)$. 
Any fixed Gaussian POVM defines a parameter-independent kernel $K_G(y|R)\ge0$ such that
\begin{equation}
p_G(y|\bm\theta)
=
\int dR\,
W_{\rho_{\bm\theta}}(R)K_G(y|R)
\label{eq:app-gaussian-kernel}
\end{equation}
Thus, every Gaussian measurement forms a classical Markov chain
$\bm\theta\rightarrow R\rightarrow Y$.

Let
$\bm S_W(R)=
\nabla_{\bm\theta}\ln W_{\rho_{\bm\theta}}(R)$
be the Wigner score.
Differentiating Eq.~\eqref{eq:app-gaussian-kernel} gives
$\bm S_G(Y)=\mathbb E[\bm S_W(R)|Y]$.
The law of total covariance then yields
\begin{equation}
\bm J_W-\bm J_G
=
\mathbb E\!\left[
\operatorname{Cov}
\bigl(\bm S_W(R)\mid Y\bigr)
\right]
\succeq0 .
\label{eq:app-gaussian-data-processing}
\end{equation}
The Schur complement minimizes the score variance after projection onto the nuisance mode. The same bound holds after eliminating the carrier phase.

Let $X_q$ be the carrier-mode amplitude quadrature.
Let
$\bm P_e=(P_{e_1},\ldots,P_{e_r})^T$,
where $r=\operatorname{rank}\bm C$, collect the phase quadratures of
the orthogonal score modes.
The effective shape score at the operating point is
\begin{equation}
\begin{aligned}
\bm S_{W,\bm\vartheta}^{\rm eff}(R)
&=
2\sqrt{2N_s}\,
\tanh\!\left(
2\sqrt{2N_s}\,X_q
\right)
\\
&\quad\times
\left(\bm C^{1/2}\right)^T\bm P_e .
\end{aligned}
\label{eq:app-effective-wigner-score}
\end{equation}
The quadrature $X_q$ determines the posterior reliability of the
unknown BPSK sign. The quadratures $\bm P_e$ carry the delay and
dispersion responses.

An ideal multimode homodyne receiver measures
$\bm Y_{\rm hom}=(X_q,\bm P_e^T)^T$.
These quadratures occupy orthogonal modes, hence commute.
Equation~\eqref{eq:app-effective-wigner-score} is fully determined by
$\bm Y_{\rm hom}$. No effective shape information remains in the
unmeasured quadratures. Therefore,
\begin{equation}
\bm J_{\bm\vartheta|\varphi_c}^{G,\max}
=
\bm J_{\bm\vartheta|\varphi_c}^{\rm hom}.
\label{eq:app-gaussian-optimum}
\end{equation}
Evaluating the covariance of
Eq.~\eqref{eq:app-effective-wigner-score} gives
Eq.~\eqref{eq:gaussian-fi-comparison} of the main text.

\section{Multimode occupied-subspace decomposition}\label{em:multimode}

Let $P$ project onto the occupied single-photon mode subspace and
$Q=I-P$. For each parameter $\vartheta_\mu$,
$g_\mu P=Pg_\mu P+Qg_\mu P$.
The first term acts within the occupied subspace. The second couples the field to unoccupied score modes. These terms then generate orthogonal local state derivatives. Eliminating the unknown common phase gives
\begin{equation}
\begin{aligned}
\bm{\mathcal K}_{\bm\vartheta|\varphi_c}^{\rm eff}
&=
\bm{\mathcal K}_{\rm leak}
+
\bm{\mathcal K}_{\rm int|\varphi_c}^{\rm eff},
\\
(\mathcal K_{\rm leak})_{\mu\nu}
&=
4\operatorname{Re}\operatorname{Tr}
\!\left[
\bm\Gamma^{(1)}
Pg_\mu Qg_\nu P
\right].
\end{aligned}
\label{eq:app-occupied-subspace-qfi}
\end{equation}
Here
$\Gamma^{(1)}_{jk}
=\langle\hat a^\dagger[u_k]\hat a[u_j]\rangle$
is the first-order coherence matrix.

For an equiprobable $M$-PSK alphabet whose symbols occupy the same mode
$q$, $P=\ket q\bra q$ and
$\bm\Gamma^{(1)}=N_s\ket q\bra q$.
Moreover,
$Pg_\mu P=\langle g_\mu\rangle_qP$.
Its second quantization is therefore proportional to $\hat N$ and is
eliminated with the common phase.
Defining
$\Delta g_\mu=g_\mu-\langle g_\mu\rangle_q$, we obtain
\begin{equation}
\begin{aligned}
\left(
\mathcal K_{\bm\vartheta|\varphi_c}^{\rm eff}
\right)_{\mu\nu}
&=
4N_s\operatorname{Re}
\bra q\Delta g_\mu\Delta g_\nu\ket q
\\
&=
4N_sC_{\mu\nu}.
\end{aligned}
\label{eq:app-rank-one-psk-qfi}
\end{equation}
The result is independent of the PSK order.
The three-mode QPG--PNRD receiver therefore remains optimal for any
such rank-one PSK alphabet.

For genuine temporal-mode multiplexing,
$P=\sum_{j=1}^{n}\ket{u_j}\bra{u_j}$.
The internal term $Pg_\mu P$ may now contain observable intermode
rotations. For mutually incoherent inputs with mean photon numbers
$N_j$, the leakage contribution is
\begin{equation}
(\mathcal K_{\rm leak})_{\mu\nu}
=
4\sum_{j=1}^{n}N_j
\operatorname{Re}
\bra{u_j}g_\mu Qg_\nu\ket{u_j}.
\label{eq:app-multimode-leak-qfi}
\end{equation}
Parallel QPG channels can measure these contributions
\cite{brecht2015photon}.
If occupied modes couple to common score modes, a joint multimode
transformation is generally required.

\section{SQL and Helstrom benchmarks}\label{em:benchmarks}

The Hadamard transformation maps the two-mode QPSK code onto two
binary branches of mean energy $N_s/2$.
The SQL follows from two homodyne decisions.
The conditional Helstrom success probability is the product of the
corresponding binary optima.

Define
$z(\eta)=
\sqrt{\eta\eta_dN_s/(1+2\eta_dN_t)}$.
Averaging over the transmittance distribution gives
\begin{equation}
\begin{aligned}
\bar P_e^{\rm SQL}
&=
\int_0^1d\eta\,p(\eta)
\Bigl[
\operatorname{erfc}z(\eta)
\\[-1mm]
&\hspace{25mm}
-\frac14
\operatorname{erfc}^{2}z(\eta)
\Bigr],
\\
\bar P_e^{\rm Hel}
&=
\int_0^1d\eta\,p(\eta)
\Biggl\{
1-\frac14
\\[-1mm]
&\hspace{14mm}\times
\left[
1+\sqrt{1-e^{-2\eta N_s}}
\right]^2
\Biggr\}.
\end{aligned}
\label{eq:app-qpsk-benchmarks}
\end{equation}
The Helstrom expression refers to the optical state before receiver
noise. It also assumes knowledge of the instantaneous transmittance.
It therefore provides an oracle-CSI lower bound.
The ideal-noise SQL follows from $\eta_d=1$ and $N_t=0$.

\section{MAP decision and coupled thresholds}\label{em:MAP}

Without instantaneous channel information, the two PNRD counts must
be processed jointly.
The receiver uses the marginalized MAP likelihood defined in the main
text. In the Poisson limit $N_t=0$, this rule has a single threshold in
either count when the other count is fixed.

Let $N_b=N_s/2$ be the energy in each branch.
For fixed $n_-$, define
\begin{equation}
\begin{aligned}
A_\pm(n_+;n_-)
&=
\max_{b_-\in\{\pm1\}}
\pi_{(\pm1,b_-)}
\\
&\quad\times
\bar p(n_+,n_-\mid\pm1,b_-).
\end{aligned}
\label{eq:app-map-scores}
\end{equation}
The MAP rule selects $b_+=+1$ when
$A_+(n_+;n_-)\ge A_-(n_+;n_-)$.

The conditional Poisson means in the plus branch are
$\mu_\pm(\eta)
=\eta_d[d_+\pm\sqrt{\eta N_b}]^2$.
If $d_+\ge\sqrt{N_b}/2$, then
$\mu_+(\eta)\ge\mu_-(\eta')$
for all $\eta,\eta'\in[0,1]$.
For $\mu_+\ge\mu_-$,
\begin{equation}
\begin{aligned}
&p_{\rm P}(n+1,\mu_+)p_{\rm P}(n,\mu_-)
\\
&\quad-
p_{\rm P}(n,\mu_+)p_{\rm P}(n+1,\mu_-)
\\
&=
\frac{\mu_+-\mu_-}{n+1}
\\
&\quad\times
p_{\rm P}(n,\mu_+)p_{\rm P}(n,\mu_-)
\ge0 .
\end{aligned}
\label{eq:app-poisson-single-crossing}
\end{equation}
Thus, the likelihood advantage of the larger-mean hypothesis is
nondecreasing with the observed count.
Averaging over fading preserves this ordering.
Maximization over the other branch preserves it as well.

The plus-branch threshold is therefore
\begin{equation}
\begin{aligned}
n_{\rm th}^{+}(n_-)
&=
\min\Bigl\{
n_+\in\mathbb N_0:
\\[-1mm]
&\hspace{12mm}
A_+(n_+;n_-)
\ge
A_-(n_+;n_-)
\Bigr\},
\\
\hat b_+=+1
&\quad\Longleftrightarrow\quad
n_+\ge n_{\rm th}^{+}(n_-).
\end{aligned}
\label{eq:app-coupled-map-threshold}
\end{equation}
The minus branch obeys the analogous rule with
$n_{\rm th}^{-}(n_+)$.
The MAP likelihood fixes both thresholds, which can be stored as lookup tables.
For finite $N_t$, the receiver evaluates the displaced-thermal MAP
scores directly. The simulations do not use the Poisson-threshold
approximation.
\end{document}